\documentclass[%
 aip,
 amsmath,amssymb,
 reprint,%
]{revtex4-1}

\usepackage{graphicx}
\usepackage{dcolumn}
\usepackage{bm}
\usepackage{color}
\usepackage{dsfont}

\usepackage[utf8]{inputenc}
\usepackage[T1]{fontenc}
\usepackage{newtxmath}
\usepackage{etoolbox}

\usepackage[normalem]{ulem}

\makeatletter
\def\@email#1#2{%
 \endgroup
 \patchcmd{\titleblock@produce}
  {\frontmatter@RRAPformat}
  {\frontmatter@RRAPformat{\produce@RRAP{*#1\href{mailto:#2}{#2}}}\frontmatter@RRAPformat}
  {}{}
}%
\makeatother
\begin{document}

\preprint{AIP/123-QED}

\title{Symmetry-breaking mechanism for the formation of {cluster} chimera patterns}

\author{Malbor Asllani}
\affiliation{School of Mathematics and Statistics, University College Dublin, Belfield, Dublin 4, Ireland}
\affiliation{MACSI, Department of Mathematics and Statistics, University of Limerick, Limerick V94 T9PX, Ireland}
\author{Bram A. Siebert}
\affiliation{MACSI, Department of Mathematics and Statistics, University of Limerick, Limerick V94 T9PX, Ireland}
\author{Alex Arenas}
\affiliation{Departament d’Enginyeria Informàtica i Matemàtiques, Universitat Rovira i Virgili, 43007 Tarragona, Catalonia, Spain}
\author{James P. Gleeson}
\affiliation{MACSI, Department of Mathematics and Statistics, University of Limerick, Limerick V94 T9PX, Ireland}

\begin{abstract}
{The emergence of order in 
collective dynamics is a fascinating phenomenon that characterizes many natural systems consisting of coupled entities.}  
{Synchronization is such an example where individuals, usually represented by either linear or nonlinear oscillators, can spontaneously act coherently with each other when the interactions' configuration fulfills certain conditions. However, synchronization is not always perfect, {and} 
the coexistence of coherent and incoherent oscillators, broadly known in {the literature as} chimera states, is also possible. Although several attempts have been made to explain how chimera states are created, their emergence, stability, and robustness remain a long-debated question. We propose an approach that aims to establish a robust mechanism through which chimeras originate. We first introduce a stability-breaking method where clusters of synchronized oscillators can emerge. Similarly, one or more clusters of oscillators may remain incoherent within yielding a particular class of patterns that we here name \emph{cluster} chimera states.}
\end{abstract}

\maketitle

\begin{quotation}
Synchronization of the dynamics of the individual entities that constitute complex systems has been observed and systematically studied for {over four decades} \cite{sync_book,arenas, newman_book}. An exotic behavior that has triggered scientists' curiosity for a long time is the emergence of chimera states, long-lasting dynamical patterns of coexistence of synchronous and asynchronous clusters of nodes in networked systems \cite{kuramoto, chimera, solvable_chimera, transient_chimera}. Although many features {are now understood}, further  questions remain unanswered \cite{panaggio,editorial_chimera}. In particular, the stability of such states for finite networks of oscillators and their high sensitivity to the initial conditions have not yet been tackled exhaustively.
{Several recipes have been suggested recently that {aim to reproduce} chimera states by decomposing the network into stable and unstable clusters \cite{pecora, sorrentino1, sorrentino2, motterPRL, motter_strong}}.
{Here we show} that the formation of clustered synchronization \cite{sync_basin, cluster_sync} can follow a global symmetry-breaking mechanism, which is also responsible for selecting the nature of the final patterns \cite{prigogine, cross, murray}. {In this way}, we {demonstrate} that it is possible to reconcile the pattern formation process with cluster synchronization, paving the way to a robust mechanism for explaining the formation of chimera states in networks of coupled dynamical systems. 
\end{quotation}


{\section{Introduction}}

The full understanding of the dynamics of complex systems is probably one of the biggest challenge of physics this century. The emergence of collective behaviour and the coexistence of order and disorder are fundamental key players in this process.
A very well--known phenomenon of collective behavior is that of synchronization, where a set of
{coupled oscillators spontaneously synchronize their phases}. This phenomenon has been widely studied in recent years and has been applied to explain the coordination of the beats of the heart myocytes and the firing of brain neurons \cite{sync_book, heart, brain_chimera}, the simultaneous flashing of male fireflies during the mating season \cite{buck} and the synchronization of the alternating current in power grids \cite{power_grid}. Theoretical studies have identified particular types of synchronization, among which it is worth mentioning the remote synchronization of two or more oscillators through asynchronous ones \cite{nicosia}, the synchronization of chaotic oscillators and the phenomenon of  cluster synchronization when the set of oscillators is divided into subsets of synchronized ones \cite{cluster_sync} or the emergence of exotic states where identically coupled oscillators synchronise in clusters \cite{raissa}. 

The latter class also includes an essential and  fascinating phenomenon {of coexistence of order and disorder} known as chimera states. Chimeras, named from Greek mythology, are particular states where clusters of coherent and incoherent oscillators coexist. Since their discovery, chimera states have aroused scientists' curiosity about the mechanisms that give birth to them, about the effective lifespan of these states \footnote{Usually by \textit{state} in the dynamical systems theory, it is referred to as an equilibrium solution. In this sense, the long transient of the chimera behavior makes it effectively a state, although its stability might not be truly asymptotic.}, and their sensitivity to the initial conditions that make their experimental verification difficult to achieve \cite{kuramoto, panaggio}. 

{Outstanding} theoretical results have been obtained {in the study of} chimeras, and promising methods have been proposed to reveal the mystery surrounding these exotic states, {such as whether they will exist, their stability, and specially, their origin}. {For example, it has been found} that chimeras are stable in infinite-size networks and that they can be transient in finite ones \cite{transient_chimera}. So far, however, there is not an exhaustive answer to the stability of chimera states \cite{chimera, solvable_chimera}, or a robust mechanism that explains how chimeras are generated \cite{motterPRL}. 
{An important advance in the area came from the group-theoretical characterization of network symmetries that corresponds to the identification of partitions of nodes that can be stable or unstable \cite{pecora, sorrentino1, sorrentino2, motterPRL}. Using this approach the authors are able to design the emergence of chimera states.}
The most recent approach to understanding the coexistence of both coherent and incoherent oscillators has been that of group-theoretical characterization of network symmetries that,  corresponds to the identification of partitions of {set} of nodes that can be stable or unstable \cite{pecora, sorrentino1, sorrentino2}. {Other recent alternatives have been based on symmetry-breaking principles where chimeras emerge from an earlier uniformly synchronized regime \cite{motterPRL, motter_strong}.}
{Aside from theoretical results,} experimental {evidence} \cite{showalter} proves the existence of cluster synchronization and chimera states for the case of coupled Belousov-Zhabotinsky oscillators \cite{prigogine}. 

{In this paper}, we aim {to analyze} the chimera states, proposing an alternative path to the aforementioned cluster partition. More explicitly, we will use a pattern formation approach based on a global symmetry-breaking mechanism \cite{prigogine, cross, murray} that yields cluster synchronization patterns as a result. {Based} on this principle, we will successively generate {coexisting clusters of coherent and incoherent oscillations that are robust to initial conditions. We name these states \emph{cluster} chimera patterns.} The great advantage of this method is that it has been extensively studied {since} the early $1970$'s, and many groundbreaking rigorous analytical results have been obtained regarding the prediction of the shape and stability of the final nonlinear patterns. {In fact}, the pattern formation process is robust to the choice of the initial conditions, {at odds with classical chimeras that are highly sensitive to the initial setting of variables \cite{kuramoto, chimera, panaggio}. Furthermore, it has been proven that patterns obtained following an instability principle are not transient, but a stability region of parameters (also known as ``{stability} balloon'') where they are asymptotically stable exists \cite{murray, cross}}.

{A major difference in our approach is the model setting through which oscillatory behavior emerges. Traditionally, the synchronization phenomenon has been studied in systems of coupled oscillators. The main reason is that global coherence is regarded as the emergent behavior in these systems. Thus it is expected that the individual nodes have to oscillate independently before fully coherent or chimera states eventually emerge. At variance with this framework, here we show that cluster synchronization and chimera patterns can also arise when the individual nodes do not manifest any oscillatory behavior. As we will next show in detail, our method is fully based on a global instability mechanism where the oscillations are also generated due to either inter-node coupling \cite{natcomm} or multispecies interactions \cite{LNA}, further facilitating the formation of synchronized or chimera clusters.}
{In fact, a key aspect of our analysis is the construction of networks whose structural properties yield the segregation of the entries of the eigenvectors of the coupling operator into clusters. 
We demonstrate  that sufficiently modular networks, in particular, possess this property.}

The {aforementioned characterizing features of the pattern formation process} are crucial because they will pave the way and facilitate obtaining chimera states. Thus, instead of tackling the problem frontally to understand how to decompose the chimera in groups of locally stable and unstable nodes, we {propose a recipe for} rigorously proving the emergence of simultaneously coherent and incoherent {oscillatory nodes}.\\

{\section{Symmetry-breaking mechanism}}

We start by considering the general setting of a network of coupled dynamical systems through the equations
\begin{equation}
    \dot{\mathbf{x}}_i=\mathbf{F}\big(\mathbf{x}_i\left(t\right)\big) + \boldsymbol{\sigma} \odot \sum_{i=1}^N W_{ij}\mathbf{G}\big(\mathbf{x}_j\left(t\right)\big),
    \label{eq:general}
\end{equation}
where each of the $N$ individuals, identified as nodes of the network, are represented by $M$-dimensional vector state variables $\textbf{x}_i$~\cite{arenas}. The local dynamics of the $i$-th node is given by the nonlinear function $\textbf{F}\left(\textbf{x}_i\right)$, $\boldsymbol{\sigma}$ is the vector of the coupling strength, $W_{ij}$ are the weighted entries of the adjacency matrix and $\textbf{G}\left(\textbf{x}_j\right)$ is the coupling function. Notice that in the above equation, $\odot$ represents the element-wise product, and we have generalised the global coupling term $\boldsymbol{\sigma}$ into a vector, a generalisation that will prove useful in the following. We start by considering the same synchronised equilibrium state for all the nodes $\textbf{x}^*(t)$, which is stable in the absence of coupling. Then by perturbing this state by a small random term $\textbf{x}_i(t)=\textbf{x}^*(t)+\boldsymbol{\xi}_i(t)$ we can obtain at the first perturbative order the variational equations $\dot{\boldsymbol{\xi}}_i=\mathbf{DF}\big(\mathbf{x}^*\left(t\right)\big)\boldsymbol{\xi}_i\left(t\right) + \boldsymbol{\sigma} \odot \sum_{i=1}^N W_{ij}\mathbf{DG}\big(\mathbf{x}^*\left(t\right)\big)\boldsymbol{\xi}_j\left(t\right)$.
Now starting from here and assuming that the linearized spatial operator $\mathbf{DG}\big(\mathbf{x}^*\left(t\right)\big)$ is diagonalizable, we can proceed further to obtain the Master Stability Function (MSF) \cite{arenas, master_stability}, which relates the stability of the uniformly synchronized state $\textbf{x}^*$ to the topology of the connection network. So, assuming that there is a linearly independent basis of eigenfunctions(-vectors) of the spatial operator, we can obtain the diagonalized decoupled equations
\begin{equation}
    \dot{\boldsymbol{\zeta}}_\alpha=\Big(\mathbf{DF}\big(\mathbf{x}^*\left(t\right)\big)_\alpha + \Lambda^{(\alpha)}\boldsymbol{\sigma} \Big)\odot\boldsymbol{\zeta}_\alpha\left(t\right)
    \label{eq:MSF}
\end{equation}
where by $\Lambda^{(\alpha)}$ we have denoted the $\alpha$-th eigenvalue of the spatial operator. To concretize our analysis and without loss of generality, throughout this text, we will consider a diffusive coupling through the Laplacian matrix $\boldsymbol{\mathcal{L}}$ whose entries are $\mathcal{L}_{ij}=W_{ij}-k_i\delta_{ij}$ with $k_i=\sum_{j=1}^NW_{ij}$. 

The formalism described so far allows the local analysis of the linear stability of the originally uniform state. Depending on the nature of the latter, eq.~\eqref{eq:MSF} might be non-autonomous when the uniform state is a limit cycle or autonomous when it is a fixed point. The traditional approach  in the study of  synchronization phenomenon is based on the former case when the system is already in a synchronized limit cycle manifold. In this case, the aim is to keep as stable as possible such a state since an eventual instability would cause the desynchronization of the whole system. {As already anticipated}, we will adopt here an alternative approach that to the best of our knowledge has not been used so far in the vast literature related to the synchronization {dynamics}. The first major difference is that we start from an unstable uniform equilibrium of a fixed point, so $\textbf{x}^*_i(t)=\textbf{x}^*$ for all the nodes. 
Starting from a uniform stationary state will allow the use of the spectral analysis instead of the classical Lyapunov exponent used for non-autonomous systems \cite{arenas, master_stability}. This makes possible the necessary analytical treatment for finding the instability conditions and the description of the orbits' {behavior} in the initial linear regime. {In the latter setting the MSF formalism is known as the \emph{dispersion relation} \cite{murray}.} Since a homogeneous stationary equilibrium lacks the oscillatory behaviour that characterises synchronization dynamics, we will recover final time-varying patterns by considering a nonlinear function constituted by $3$ variables, known in the {biomathematical} literature as species~\cite{murray} $\textbf{F}\left(\textbf{x}_i\right)=\big[f\left(x_i\right), g\left(y_i\right), h\left(z_i\right)\big]$. For each node we write the $3$ dimensional state vector $\textbf{x}_i=\left[u_i, v_i, z_i\right]$ with a little abuse of notation. In other words, we have            
\begin{equation}\label{eq:3species}
    \begin{cases}
        \dot{u}_i = f(u_i,v_i,z_i) + \sigma_u \sum\limits_{j=1}^N \boldsymbol{\mathcal{L}}_{ij}u_j\\[.25cm] 
        \dot{v}_i = g(u_i,v_i,z_i) + \sigma_v \sum\limits_{j=1}^N \boldsymbol{\mathcal{L}}_{ij}v_j\\[.25cm] 
        \dot{z}_i = h(u_i,v_i,z_i) + \sigma_z \sum\limits_{j=1}^N \boldsymbol{\mathcal{L}}_{ij}z_j.
    \end{cases}
\end{equation}

\begin{figure*}[t!]
    \centering
    \includegraphics[width = 1\linewidth]{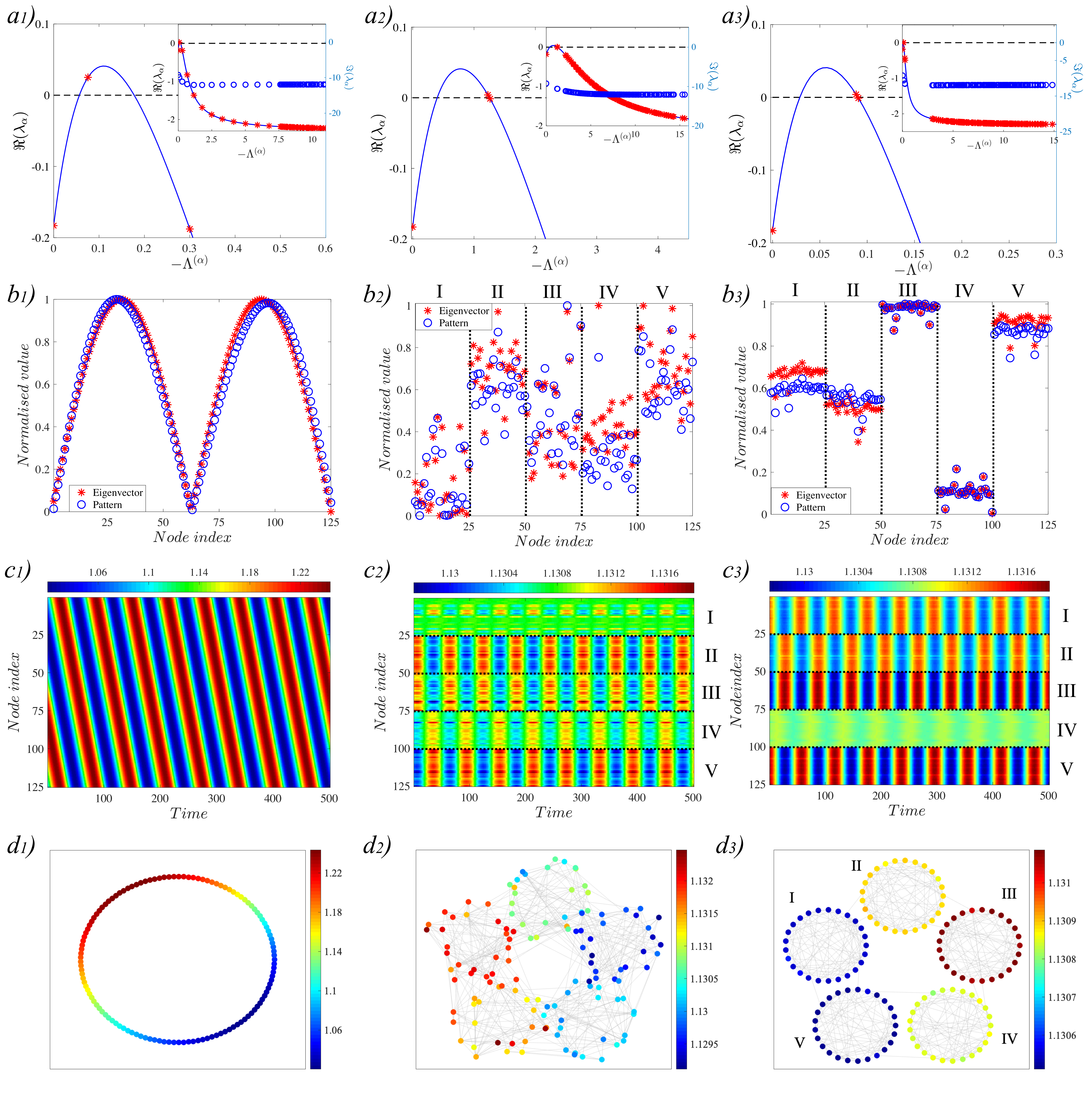}
    \caption{\textbf{Cluster synchronisation in modular networks} The $3$ species model used to generate the patterns follows that of Zhabotinsky \cite{zhabotinsky, LNA} defined as $f(u,v,z)=-c_1uv^2+c_3-c_7u/(q+u)$ with $q=0.0001$, $g(u,v,z)=c_2uv^2-c_4v+c_8$, and $h(u,v,z)=c_5u-c_6v$ with general parameters $c_1=c_3=28.2$, $c_2=c_4=15.5$, $c_5=c_6=1$, $c_7=25.38$, $c_8=3.1$, $\sigma_u=\sigma_v=0$ and only $\sigma_z$ is varying according to the specific case. In the columns from left to right are shown the networks in increasing modularity, from a ring lattice (left) with $\sigma_z=30$ to a strong modular one (right) $\sigma_z=59$ passing through a intermediate modularity (center) $\sigma_z=4.25$.  $\mathbf{a_1)-a_3)}$ The dispersion relation for the different setting of clusterisation of the network where a single unstable mode has been selected. In the inset are shown both the real and the imaginary part of the dispersion relation. $\mathbf{b_1)-b_3)}$ Comparison between the eigenvectors of the corresponding unstable modes vs. the final pattern. Notice the change in the shape of the eigenvector from a Fourier (discretised) eigenfunction to a cluster segregated one. $\mathbf{c_1)-c_3)}$ Time evolution of patterns {of the first species}, from a travelling wave in the ring lattice to a cluster synchronised one for the strong modular network. {Also the small size of the pattern is due to the tiny value of real part of the unstable eigenvalue, purposely selected near the bifurcation threshold. ({A} larger pattern {is} found in {Appendix \ref{app:B}}.} $\mathbf{d_1)-d_3)}$ Graphical representation of snapshots of the oscillatory patterns.} 
    \label{fig:sync}
\end{figure*}

It is well known that a $3$ species system is a sufficient requirement to have an oscillatory instability and, as a consequence, nonlinear oscillatory patterns \cite{cross,murray}. This is the key factor in producing the global oscillations of our system, as is shown in Fig. \ref{fig:sync} panels {$a_1)-a_3)$}. As an alternative to this model, oscillatory patterns can be obtained also for a $2$ species model in a directed network where the oscillations are as a consequence of the complex spectrum of the Laplacian matrix \cite{natcomm, epjb} (see  {Appendix \ref{app:B}} for a detailed description). {When neither of the two aspects above contributing to oscillations occurs, then the instability is stationary, and modular Turing patterns can emerge \cite{turing, bram}.} {We should, however, emphasise that in general, in a complex network, the oscillations appear to be asynchronous following the shape of a traveling front in the limit case of a regular lattice~\cite{cross,othmer1} (see Fig. \ref{fig:sync} {$c_1)$} and {$d_1)$)}.} 
To understand the rationale behind such behavior, we should look for the shape of the {eigenvectors}, which describe the linear evolution of the orbits in a short time and also {determine} the general shape of the final nonlinear pattern. In fact in order to solve the linear (decoupled) variational equations \eqref{eq:MSF} we look for solutions of the form $\boldsymbol{\zeta}_i=\sum_{\alpha=1}^N c_\alpha e^{\gamma_\alpha}\Phi^{(\alpha)}_i$ where $\gamma_\alpha$ and $\Phi^{(\alpha)}$ represent the $\alpha$-th growth mode and Laplacian eigenvector respectively. So, in the initial regime, once there is at least one value of $ \alpha $ for which the real part of $\gamma_\alpha$, the dispersion relation {defined earlier}, is positive, as is the case for Fig. \ref{fig:sync} $a_1)$, then the entries of vectors of the $3$ species will be scaled by the entries of the eigenvector $\Phi^{(\alpha)}$. Also, in a $3$ species model, the imaginary part is different from zero allowing an oscillatory instability to occur \cite{cross}. In regular multi-dimensional lattices, such an eigenvector represents a discrete term of multi-dimensional Fourier series. This is why in such discrete domains, the final patterns will be a combination, depending on the number of the unstable (Fourier) modes, of time-varying sinusoidal shaped traveling waves \cite{cross, murray,othmer1}.


{\subsection{Network construction for clustered eigenvectors}}

Having briefly introduced the mechanism responsible for shaping the stable nonlinear pattern, we next describe how to construct the coupling network to obtain clusters of coherent and incoherent nodes \footnote{We want to emphasize here that the concept of nodes \textit{being in synchrony} with each other is, of course, relativistic. Originally Kuramoto introduced in his famous model an order parameter to quantify the amount of synchronization \cite{kuramoto}.}. {We} start considering the set of eigenvectors $\left[\Phi^{(0)}, \Phi^{(1)},\dots,\Phi^{(N)}\right]$ of the Laplacian of a connected and undirected network sorted in decreasing order of the respective modes (eigenvalues). From the algebraic connectivity, we know that the network Laplacian will have a single zero eigenvalue $\Lambda^{(0)}=0$ whose eigenvector is uniformly distributed, for simplicity here assumed to be the scaled unit vector $\Phi^{(0)}=a\mathds{1}=\left[a, a, \dots, a\right]^T$ where $a$ is the scaling constant. Let us remark here that from eq. \eqref{eq:MSF} the null mode $\Lambda^{(0)}=0$ represents the a-spatial case meaning that the nodes should be stable once uncoupled. So we should look for the eigenvectors corresponding to the nonnull modes to destabilize {our system globally}. 
To obtain eigenvectors with clusters of entries that are close to each other, in total analogy with the equilibrium state, we proceed by construction. We start by considering networks with more than one null mode, for example, $C$ such eigenvalues ${}_1\Lambda^{(0)}=\dots={}_C\Lambda^{(0)}=0$. This means that our network has $C$ disconnected components, here referring to as disconnected clusters. The Laplacian matrix and its eigenvectors in this case are respectively a block matrix and block vectors
    $\boldsymbol{\mathcal{L}}=diag\left[
    {}_1\boldsymbol{\mathcal{L}},\dots,{}_n\boldsymbol{\mathcal{L}},\dots, {}_C\boldsymbol{\mathcal{L}}\right]$ and
    $\Phi^{(\alpha)}=\left[{}_1\Phi^{(\alpha)} ,\dots, {}_n\Phi^{(\alpha)} ,\dots, {}_C \Phi^{(\alpha)}\right]$.
Here, we have divided the $N$ nodes into $C$ disconnected clusters with the same number of nodes. Observe that the eigenvector of the $n$-th null mode can include different solutions, but here we chose purposely the one that has constant entries for the blocks of each cluster
$${}_n\Phi^{(0)}=\big[\underbrace{{}_na_1\dots, {}_na_1}_{N/C}\dots,\underbrace{{}_na_i \dots, {}_na_i}_{N/C}\dots\underbrace{{}_na_C\dots, {}_na_C}_{N/C}\big].$$ Instead for the eigenvectors of the non null modes the only possible solution is $${}_n\Phi^{(\alpha\neq 0)}=\big[\underbrace{0\dots, 0}_{N/C}\dots,\underbrace{{}_n\Phi^{(\alpha)}}_{N/C}\dots\underbrace{0\dots, 0}_{N/C}\big]$$ where ${}_n\Phi^{(\alpha)}$  is the $\alpha$-th eigenvector of the $n$-th disconnected cluster. We invite the interested reader to refer to  {Appendix \ref{app:A}} for a rigorous derivation of this assertion.
For the next step, we weakly connect the $C$ disconnected clusters through a minimal number of links. From a purely perturbative point of view, this means that the former spectrum will be weakly deformed. The first indication of this deformation is that due to the connectedness, the graph Laplacian will have only one zero eigenvalue. All the remaining  $C-1$  formerly zero eigenvalues will be different from zero, but very small in value. Also, in terms of equilibrium states, now we have the only one eigenvector with uniformly distributed entries. So the question that arises naturally is what happened to the remaining former $C-1$ equilibrium eigenvectors? Their entries will be slightly deformed, too, so the $C-1$ eigenvectors will have almost uniform entries per block. {This essential spectral property of weakly disconnected subgraphs will be the basis for developing our method for the emergence of synchronized clusters and cluster chimera states.}\\ 

{\section{Emergence of cluster synchronized and chimera patterns}}

As a result of the above reconstruction procedure, we have obtained particular network structures that contain potential candidates of unstable modes with eigenfunctions that have entries within clusters very close to each other. Such networks indeed exist in the literature, and they are known as modular networks~\cite{newman_book}. They are widespread in many biological and social settings \cite{newman_book}, making potentially {either} the cluster synchronization {or the cluster chimera states} typical emergent behaviors in a wide class of networks. In Fig. \ref{fig:sync} we show how the oscillations in a $3$ species reaction-diffusion system change from initially asynchronous (in the left column panels) to synchronous per cluster when the modularity increases (central and right columns panels). The pattern is organized in a traveling wave on the ring network where the modularity is absent but where the network is organized in distinct modules, the species segregate their phases following the clusters they belong to. 
\onecolumngrid
\vspace*{.5cm}
\begin{figure*}[h!]
    \centering
    \includegraphics[width = .75\linewidth]{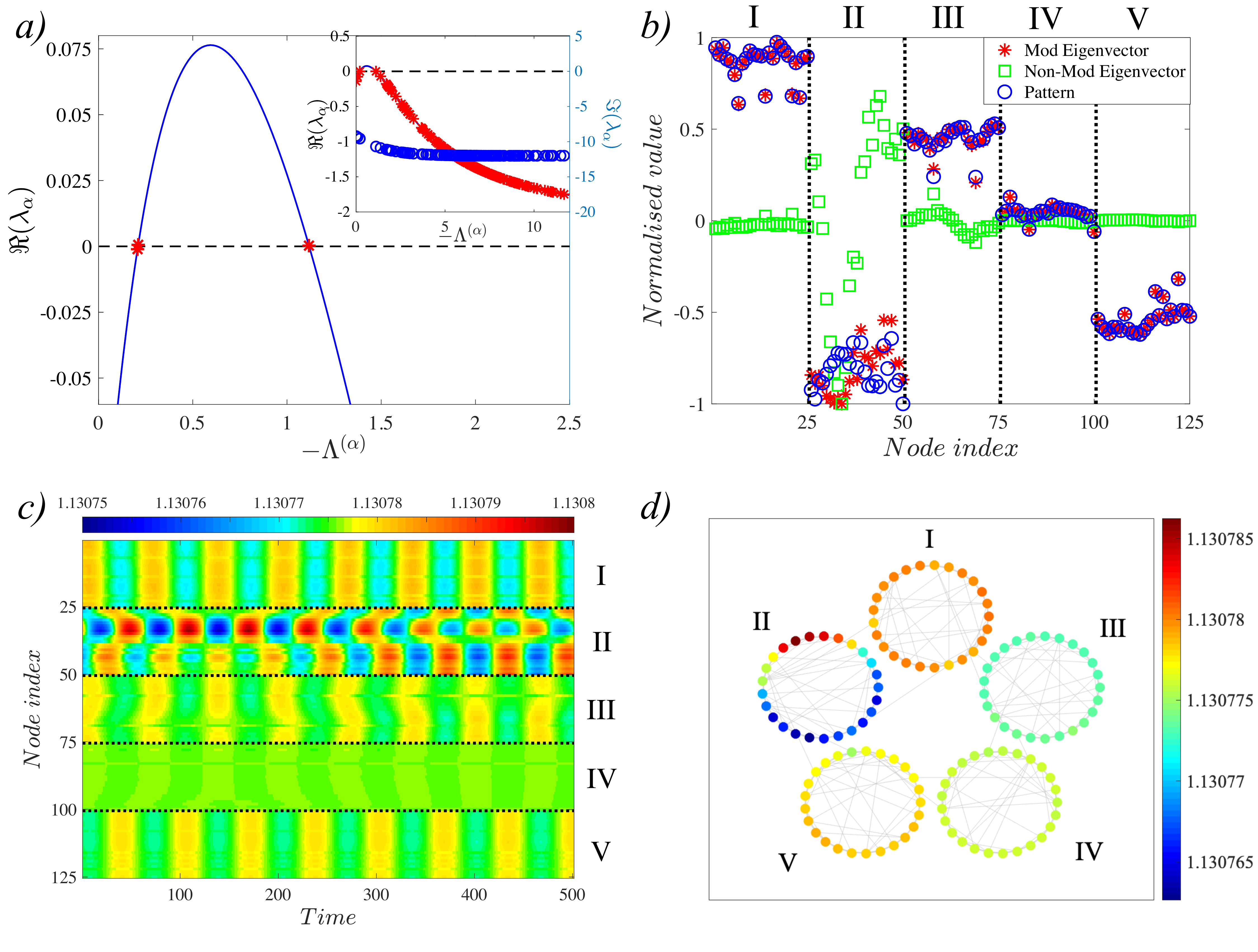}
    \caption{\textbf{Chimera states in modular networks}\,\, $\textbf{a)}$ The dispersion relation shows that in order to obtain chimera states, (at least) two modes a modular and a non-modular should be simultaneously unstable. In the inset are shown both the real and the imaginary part of the dispersion relation. $\textbf{b)}$ A comparison between the eigenvectors of the selected unstable modular (red stars) and non-modular (green squares) eigenvalues are shown here in comparison with the final pattern (blue circles) taken at a given fixed time. Notice how the irregularity of the non-modular eigenvector at the second module influences also the final pattern. \textbf{c)} The chimera states evolution {of the first species} is shown as a function of time where the second module is incoherent compared to all the remaining four ones. $\textbf{d)}$ A graphical representation of the snapshots of the chimera states in the $5$ modules network. The model is the same as in Fig. \ref{fig:sync} and the parameters used in this case are $c_1=c_3=27.88$, $c_2=c_4=15.5$, $c_5=c_6=1$ $c_7=25.092$, $c_8=3.1$  $\sigma_u=\sigma_v=0$ and $\sigma_z=5.5$. }
    \label{fig:chimera}
\end{figure*}

\twocolumngrid
\noindent
The explanation for that can be found in the eigenvectors of the first $C-1$ non-zero eigenvalues, {previously introduced}, that we denote here as \emph{modular} eigenvalues. 
The first of them that we have \sout{also} selected to be unstable in Fig. \ref{fig:sync} $a_1)$ is known as the Fiedler eigenvalue, and it is an indicator of the algebraic connectivity of the network \cite{newman_book}.  
{As shown in the previous section, the modular eigenvectors have the unique property of being segregated per module, a feature that, when activated by the corresponding unstable modular eigenvalue, is reflected in the final nonlinear pattern.} Besides the modular eigenvalues, we also have the \emph{non-modular} ones, the remaining $N-C$ eigenvalues. Following the perturbative analysis {developed so far}, it is not possible that {non-modular} eigenvectors have segregated nonnull entries per module. 
This means that if a non-modular eigenvalue would be the only unstable mode then the pattern of such setting would be irregular for a given module and almost not expressed for the rest of them. The linear stability {principle, upon which we base our method}, allows the extension of pattern selection beyond the single unstable mode case. 
In fact, if two or more modes are unstable and near the bifurcation threshold, they will compete with each other in the linear regime until the nonlinearities stabilize the pattern shaping. {So one should} expect that the {patterning} outcome will reflect {to some extent} the shape of the eigenvectors of {all} the unstable modes {involved}. 
To show {such} behaviour, in Fig. \ref{fig:chimera} we have taken into consideration only two unstable eigenvalues, a modular and a non-modular one. The choice has been such that in the dispersion relation representation, the two modes have comparable real parts. In this way the two different behaviours will be present to similar levels. 
In fact, as can be seen from the pattern snapshot taken at {some} given time, Fig. \ref{fig:chimera} $b)$, the second module is non-homogeneous in the concentration (of the first species in this case). 
{Consequently, in} panel $c)$ {we can observe that this oscillatory pattern will} constitute a {new state that we name as a cluster} chimera. {Let us note that the choice for the second module to be incoherent is simply due to the intra- and inter-connections, but other scenarios where different or more incoherent modules are also possible, as can be seen in Appendix \ref{app:B}. Although} the results presented in both {previous} figures have been validated for the case of modules of the same size, in the {following} we show that our method is robust even for the case of networks with modules of considerably different {in size.}

{\section{Robustness of the symmetry-breaking method for modules of different sizes}}

\begin{figure}[h!]
    \centering
    \includegraphics[width = 1\linewidth]{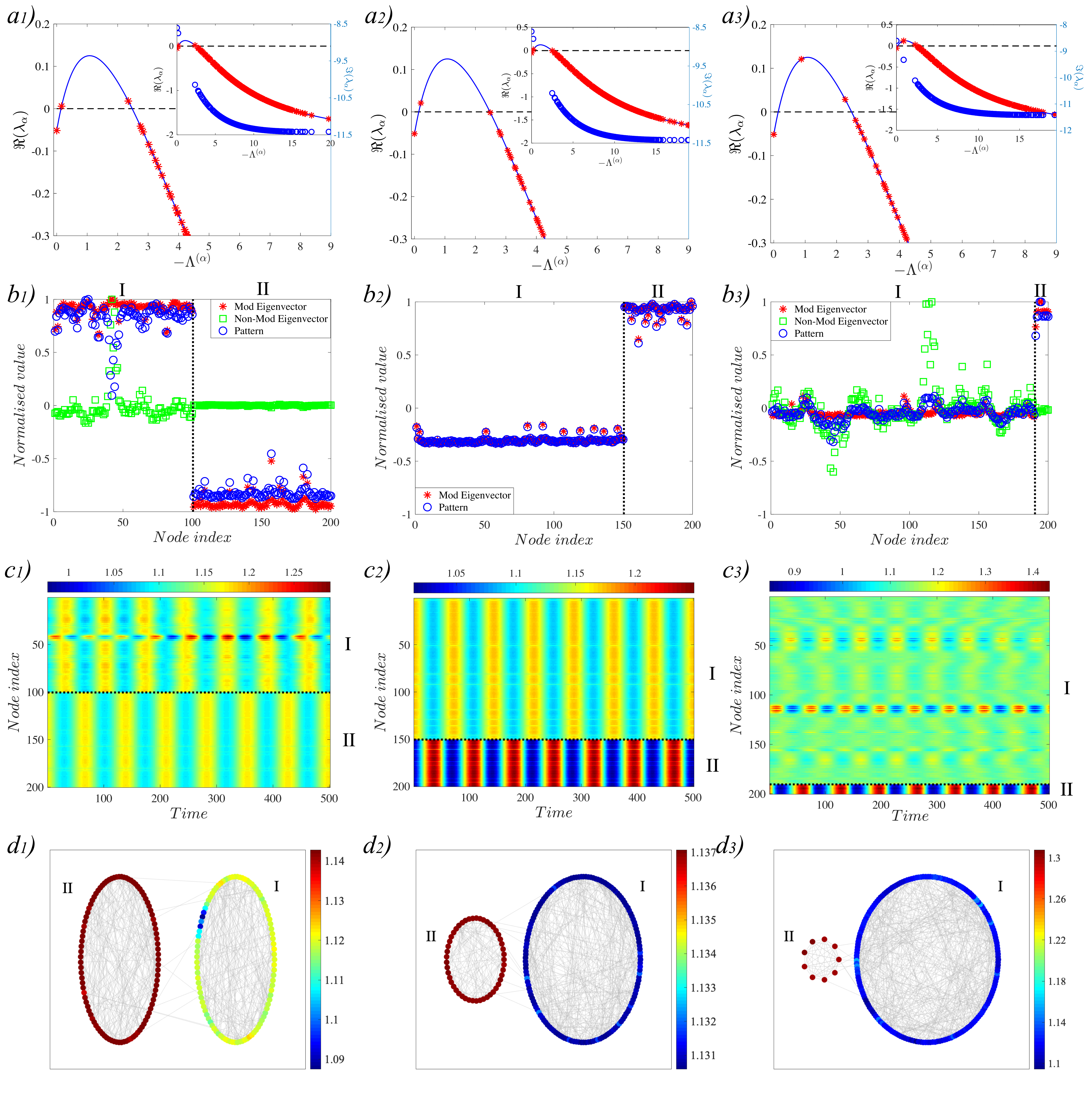}
    \caption{{\textbf{Robustness of the symmetry-breaking mechanism for cluster synchronisation and chimera states in modular networks} We have used the Zhabotinsky model as defined in the main text, with parameters $c_1=c_3=27$, $c_2=c_4=14.5$, $c_5=c_6=1$, $c_7=24.3$, $c_8=2.9$, $\sigma_u=\sigma_v=0$, and $\sigma_z=2.6$. In the columns from left to right are shown the networks in increasing ratio between the modules, respectively from a $100$ vs. $100$ for left panels to a $150$ vs. $50$, middle panels, and concluding with $190$ vs. $10$, right panels.  $\mathbf{a_1)-a_3)}$ The dispersion relation for the modules of different sizes. In the inset are shown both the real and the imaginary part of the dispersion relation. $\mathbf{b_1)-b_3)}$ Comparison between the eigenvectors of the corresponding unstable modes vs. the final pattern. $\mathbf{c_1)-c_3)}$ Time evolution of patterns, from a weakly chimera pattern to a cluster synchronized one. $\mathbf{d_1)-d_3)}$ Graphical representation of snapshots of the oscillatory patterns.}} 
    \label{fig:2modules}
\end{figure}

The method that we proposed {throughout this paper} has been {successful} in the generation and prediction of the cluster synchronization, {and the} {cluster} chimera states.
{In this} section will discuss the {robustness} of the symmetry-breaking mechanism when the modules' size is considerably different. 
We start by considering a simple network model of $NW$ topology of $200$ nodes divided into two modules with a number of inter-edges between the modules few on average compared to the intra-edges ones. 
Keeping a small {number of inter-edges compared to the number of intra-edges} is a necessary requisite for the validity of the perturbative approach used to justify our method. 
Initially, the two modules have $50$ nodes each of as shown in panels $a_1)-a_4)$ {of} Fig. \ref{fig:2modules}, and we chose the parameters in such a way to obtain a {(weak) cluster} chimera where module I is weakly incoherent due to a mix of unstable modular and non-modular eigenvectors. 
Clearly, in this setting, the results are shown throughout this work stand firmly, as confirmed by the clear gap between the modular and non-modular set of eigenvalues, panel $a_1)$. 
Then we tune the network topology by keeping the same total number of links but changing the size of the modules in the $150$ vs. $50$ ratio. 
We notice that the first two non-zero eigenvalues' {position} slightly shifts to the right, panel $a_2)$, so that the modular eigenvalue increases the spectral gap from the origin {yielding a cluster synchronized pattern}. 
When we raise the ratio to $190$ vs. $10$ nodes{, respectively,} per module, we can observe a neat shift of the modular eigenvalue and a decrease of the gap with the non-modular one, panel $a_3). 
${The resulting pattern is made by two synchronized clusters where one of them is slightly incoherent}. 
Such behavior is understandable from the spectral perturbative perspective. 
In fact, when we have smaller modules, i.e., of $10$ nodes as in our case, the action of the unitary weighted interconnection links (that remain invariant in number in all the three examples) over the small module can be considered less perturbative compared to the previous scenarios when such module was {larger, i.e.,} of $100$ nodes. 
Nevertheless, the results of Fig. \ref{fig:2modules}, in particular panels $a_2)$, $b_2)$, and $c_2)$ show that the shape of the eigenvectors remains still sufficiently segregated {(proportionally to the amount of modularity of the network)} to yield an explicit cluster synchronization {or chimera states} making it an obvious indicator of the robustness of our method.

{\section{Discussion and conclusions}}

To conclude, in this paper we have studied the problem of the emergence of {a new class of chimera states, which we call cluster chimeras, following a cluster synchronization mechanism based on} the modular structure of the underlying network. 
To understand the mechanism that generates these states we have developed a theoretical framework based on a symmetry-breaking principle. 
Due to a global instability the system departs from {a static} homogeneous state and the perturbations associated to each node will be shaped according to the eigenvectors corresponding to the unstable modes of the Laplacian matrix. 
This linear regime behaviour is saturated by the nonlinear terms in the final stage, inheriting the eigenvector characteristics in the spatially extended pattern. 
{As a first step, we} show that is possible to reconstruct networked structures that have particular eigenvectors with certain features: they are organised in groups of entries close to each other. 
Based in a perturbative approach, we prove that these networks have $C-1$ such eigenvectors, denoted as modular ones, where $C$ is the number of their clusters. On the other hand the remaining $N-C$ non equilibrium eigenstates, here called non-modular ones, present irregular shapes. At this point we could select (at least) two unstables eigenvalues, one modular and     
\onecolumngrid

\twocolumngrid
\noindent
one non-modular in the Master Stability Function such that due to their competition during the pattern formation process the final result {yields a cluster} chimera state. 
{On the other hand, if the unstable eigenvalues are exclusively modular ones, then the resulting pattern is made of synchronized clusters.} 

The pattern formation mechanism presents {significant advantages; first and in contrast to classical chimeras, the generation process is robust to the choice of initial conditions. Secondly, patterns that follow such a global instability principle are known to be stable and not transient, which is also observed in our numerical simulations.} 
{These two aspects are} are the main {open questions regarding the process through which chimera states emerge.} 
{As part of the novelty, in} this work we have particularly studied the case when the coupled dynamical systems are fixed points instead of the usual (nonlinear) oscillators. 
The main reason for that was because in this way it was possible to make use of the solid classical results already existing in the theory of pattern formation. 
Nevertheless we believe that the results shown here can be {also} extended to the case of coupled oscillators. 
We are confident that our results will shed light on a better understanding of the generation mechanisms and stability of chimera states and also fill the theoretical gap of a mathematical explanation for recent experimental observations in this domain \cite{showalter}.\\       

\begin{acknowledgments}
B.A.S. acknowledges funding from the Irish Research Council under Grant No. GOIPG/2018/3026. The work of J.P.G. and M.A. is partly funded by Science Foundation Ireland (Grants No. 16/IA/4470, No. 16/RC/3918 and No. 12/RC/2289 P2) and cofunded under the European Regional Development Fund. A.A. acknowledges financial support from Spanish MINECO (grant PGC2018-094754-B-C21), Generalitat de Catalunya (grant No.\ 2017SGR-896), Universitat Rovira i Virgili (grant No.\ 2019PFR-URV-B2-41), Generalitat de Catalunya ICREA Academia, and the James S. McDonnell Foundation (grant \#220020325).
\end{acknowledgments}

\appendix

{\section{Laplacian eigenvectors of $C$ weakly connected clusters network}\label{app:A}}

Starting from a perturbative setting is possible to show that the eigenvector of the $n$-th null mode {of $C$ disconnected clusters} can select only one of the possible solutions, the one that has constant entries of alternating signs for the blocks of each cluster
$${}_n\Phi^{(0)}=\big[\underbrace{{}_na_1\dots, {}_na_1}_{N/C}\dots,\underbrace{{}_na_i \dots, {}_na_i}_{N/C}\dots\underbrace{{}_na_C\dots, {}_na_C}_{N/C}\big]^T.$$ In fact, the most intuitive eigenvector of the null mode is the canonical one $${}_n\Phi^{(0)}=\big[\underbrace{0\dots 0}_{N/C}\dots,\underbrace{{}_n1 \dots {}_n1}_{N/C}\dots,\underbrace{0\dots 0}_{N/C}\big]^T.$$ However, this choice is not possible since it does not respect the perturbative approach once the clusters are weakly connected. To prove that we consider the simple case of $2$ clusters, but the same easily follows for the case of $C$ clusters. The perturbative eigenvalues problem for the canonical eigenvector is now written as 
\begin{equation*}
    \overbrace{\begin{bmatrix}
    {}_1\boldsymbol{\mathcal{L}} & \mathbf{E}_{12}\\
    \mathbf{E}_{21} & {}_2\boldsymbol{\mathcal{L}}\\ 
    \end{bmatrix}}^{\boldsymbol{\mathcal{L}}}
    \begin{bmatrix}
    \boldsymbol{\mathds{1}} +\boldsymbol{\epsilon}_1 \\
    \boldsymbol{\epsilon}_2
    \end{bmatrix}=\lambda_\epsilon \begin{bmatrix}
    \boldsymbol{\mathds{1}} +\boldsymbol{\epsilon}_1 \\
    \boldsymbol{\epsilon}_2
    \end{bmatrix}.
\end{equation*}
where $\mathbf{E}_{12}$, $\mathbf{E}_{21}$ are the intra-cluster weak coupling, $\boldsymbol{\epsilon}_1$, $\boldsymbol{\epsilon}_2$ the perturbation of the uncoupled eigenvectors and $\lambda_\epsilon$ the pertubation of the null mode. From this relation we obtain the following equalities
\begin{equation*}
    \begin{cases}
    {}_1\boldsymbol{\mathcal{L}}\boldsymbol{\epsilon}_1+\mathbf{E}_{12}\boldsymbol{\epsilon}_2=\lambda_\epsilon\boldsymbol{\mathds{1}}+\lambda_\epsilon\boldsymbol{\epsilon}_1 \;:\; \mathcal{O}(\epsilon)= \mathcal{O}(\epsilon)\\
    \mathbf{E}_{21}\boldsymbol{\mathds{1}}+\mathbf{E}_{21}\boldsymbol{\epsilon}_1+{}_2\boldsymbol{\mathcal{L}}\boldsymbol{\epsilon}_2=\lambda_\epsilon\boldsymbol{\epsilon}_2\;:\; \mathcal{O}(\epsilon)\stackrel{?}{=} \mathcal{O}(\epsilon^2)
    \end{cases}.
\end{equation*}
Clearly since the last relation is not possible this choice for the eigenvector is not permitted. 

On the contrary, for the eigenvectors of the non null modes the only possible solution is $${}_n\Phi^{(\alpha\neq 0)}=\big[\underbrace{0\dots, 0}_{N/C}\dots,\underbrace{{}_n\Phi^{(\alpha)}}_{N/C}\dots\underbrace{0\dots, 0}_{N/C}\big]^T$$ where ${}_n\Phi^{(\alpha)}$ is the $\alpha$-th eigenvector of the $n$-th disconnected cluster. In fact, if for the $2$ cluster case we consider as candidate $\big[{}_n\Phi^{(\alpha)},\mathds{1}\big]^T$, {where instead of $\mathds{1}$ we could chose any other nonzero vector,} from the same reasoning as above we have   
\begin{equation*}
    \overbrace{\begin{bmatrix}
    {}_1\boldsymbol{\mathcal{L}} & \mathbf{E}_{12}\\
    \mathbf{E}_{21} & {}_2\boldsymbol{\mathcal{L}} \\
    \end{bmatrix}}^{\boldsymbol{\mathcal{L}}}
    \begin{bmatrix}
     {}_n\Phi^{(\alpha)}+\boldsymbol{\epsilon}_1 \\
    \boldsymbol{\mathds{1}} + \boldsymbol{\epsilon}_2
    \end{bmatrix}=\left(\Lambda^{(\alpha)}+\lambda_\epsilon\right) \begin{bmatrix}
     {}_n\Phi^{(\alpha)}+\boldsymbol{\epsilon}_1 \\
    \boldsymbol{\mathds{1}} + \boldsymbol{\epsilon}_2
    \end{bmatrix},
\end{equation*}
so
\begin{equation*}
    \begin{cases}
    {}_1\boldsymbol{\mathcal{L}}\boldsymbol{\epsilon}_1+\mathbf{E}_{12}\boldsymbol{\mathds{1}}+\mathbf{E}_{12}\boldsymbol{\epsilon}_2=\lambda_\epsilon{}_n\Phi^{(\alpha)}+\Lambda^{(\alpha)}\boldsymbol{\epsilon}_1+\lambda_\epsilon\boldsymbol{\epsilon}_1 \\
    \mathbf{E}_{21}{}_n\Phi^{(\alpha)}+\mathbf{E}_{21}\boldsymbol{\epsilon}_1+{}_2\boldsymbol{\mathcal{L}}\boldsymbol{\epsilon}_2=\Lambda^{(\alpha)}\boldsymbol{\mathds{1}}+\Lambda^{(\alpha)}\boldsymbol{\epsilon}_2+\lambda_\epsilon\boldsymbol{\mathds{1}}+\lambda_\epsilon\boldsymbol{\epsilon}_2
    \end{cases}.
\end{equation*}
This is equivalent to 
\begin{equation*}
    \begin{cases}
    \mathcal{O}(\epsilon)= \mathcal{O}(\epsilon)\\
    \mathcal{O}(\epsilon)\stackrel{?}{=} \mathcal{O}(\epsilon^0)
    \end{cases}
\end{equation*}
which again cannot be true. This concludes that the only possible {choices} for the Laplacian eigenvectors of a network with disconnected clusters are defined as above.\\

{\section{Cluster synchronised and chimera states in modular directed networks}\label{app:B}}

In this section we will consider the case when the cluster synchronisation and chimera states emerge due to the modularity of directed networks. 
A major contribute of directed networks is due to the fact that the spectrum of the Laplacian matrix is complex which cause a Turing-Hopf instability in the linear regime{, known as topology-driven instability \cite{natcomm},} yielding this way oscillatory patterns. 
Let us remind here that such patterns are not otherwise possible for a $2$ species reaction-diffusion system. 
Also the directedness of the networked structure contributes in the increment of chances for emergence of patterns due to a higher dispersion relation compared to the symmetric network or the continuous domain. 
For further details the interested reader should refer to \cite{natcomm,epjb}. 
It is important also to remind that, in general, a directed networked structure might be also strongly non-normal as observed in many real-world networks \cite{NN}. 
In this case the pattern formation process follows different mechanisms induced by the non-normality of the Laplacian matrix \cite{JTB}. However, we decided to skip such scenario that goes beyond the scope of this paper and focus {on the topology-driven case only}.   

We start by considering a $2$ species activator-inhibitor model of a reaction-diffusion system. The concentration for the two species are denoted $u_i$, for the activator and $v_i$, for the inhibitor and where the index $i$ refers to one of the $N$ {nodes}. The corresponding equations take the following form:
\begin{eqnarray}
\begin{cases}
\dot{u}_i = f(u_i,v_i) +  D_u \sum\limits_{j=1}^N \mathcal{L}_{ij} u_j\\[.25cm]
\dot{v}_i = g(u_i,v_i)+ D_v \sum\limits_{j=1}^N \mathcal{L}_{ij} v_j
\end{cases}
\label{eq:RD}
\end{eqnarray}
where $D_u$ and $D_v$ are the diffusion constants and $f(\cdot, \cdot)$, $g(\cdot, \cdot)$ are the two nonlinear functions representing the interaction of the species inside each node $i$. The spatial interactions on the other side, are represented by the diffusion operator with entries $\mathcal{L}_{ij}$ which are a reflection of the topology of the networked support.

\begin{figure*}[h!]
    \centering
    \includegraphics[width = 1\linewidth]{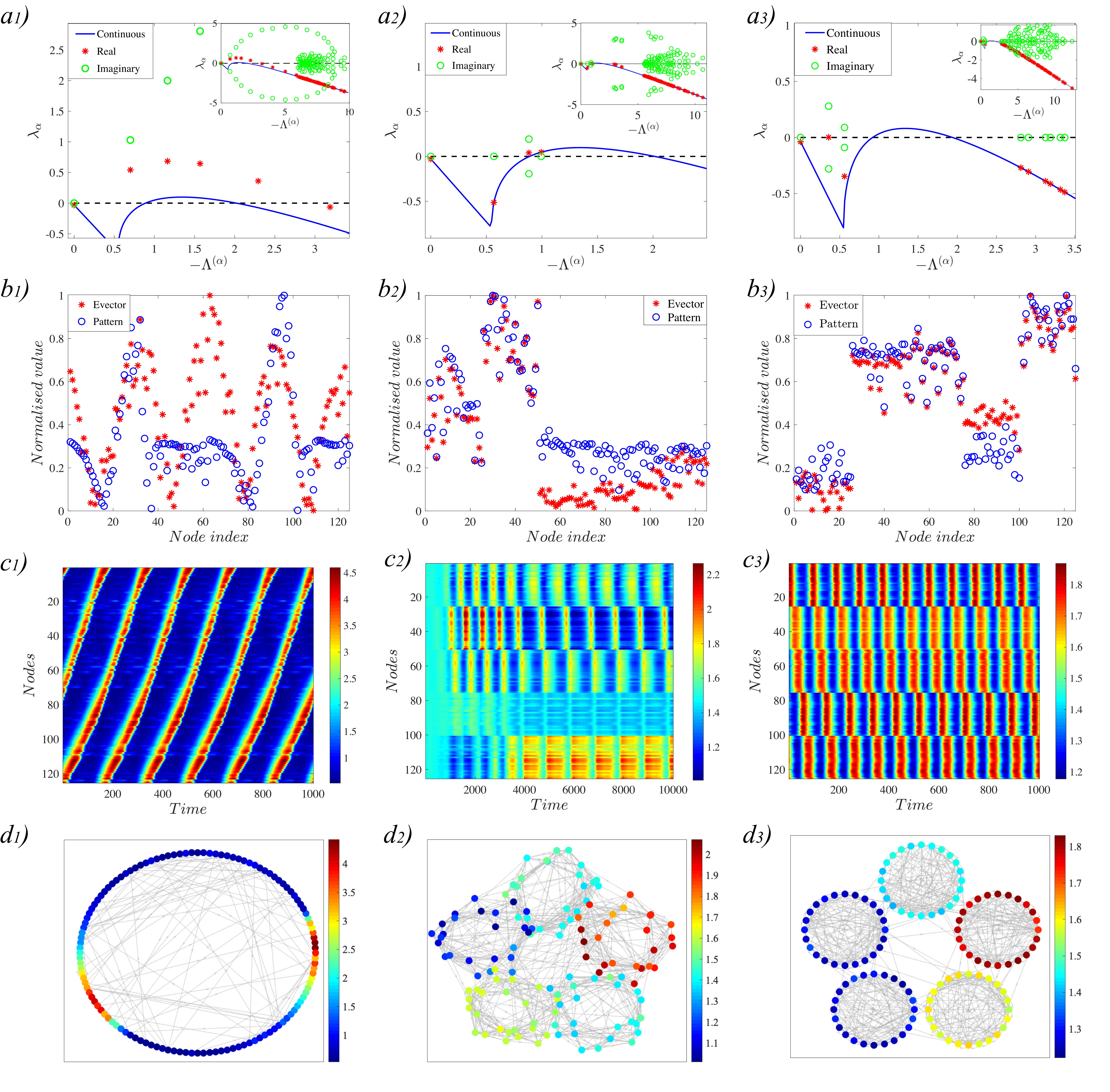}
    \caption{\textbf{Cluster synchronisation in directed modular networks} The $3$-species model used to generate the patterns follows that of Brusselator $f(u,v)=1-(b+1)u+cu^2v$ and $g(u,v)=bu-cu^2v$ \cite{prigogine} where the parameters \dots are fixed and and only $\sigma_z$ is varying according to the specific case. In the columns from left to right are shown the networks in increasing modularity, from a ring lattice (left) with $\sigma_z=30$ to a strong modular one (right) $\sigma_z=59$ passing through a intermediate modularity (center) $\sigma_z=4.25$.  \textbf{(a)} The dispersion relation for the different setting of clusterisation of the network where a single unstable mode has been selected. In the inset are shown both the real and the imaginary part of the dispersion relation. Let us notice that for a directed network the (discrete) dispersion relation is always higher then the continuous counterpart. \textbf{(b)} Comparison between the eigenvectors of the corresponding unstable modes vs. the final pattern. Notice the change in the shape of the eigenvector from a Fourier (discretised) eigenfunction to a cluster segregated one. \textbf{(c)} Patterns evolution, from a travelling wave in the ring lattice to a cluster synchronised one for the strong modular network. \textbf{(d)} Graphical representation of snapshots of the oscillatory patterns.} 
    \label{fig:direct_sync}
\end{figure*}

\begin{figure*}[h!]
    \centering
    \includegraphics[width = 1\linewidth]{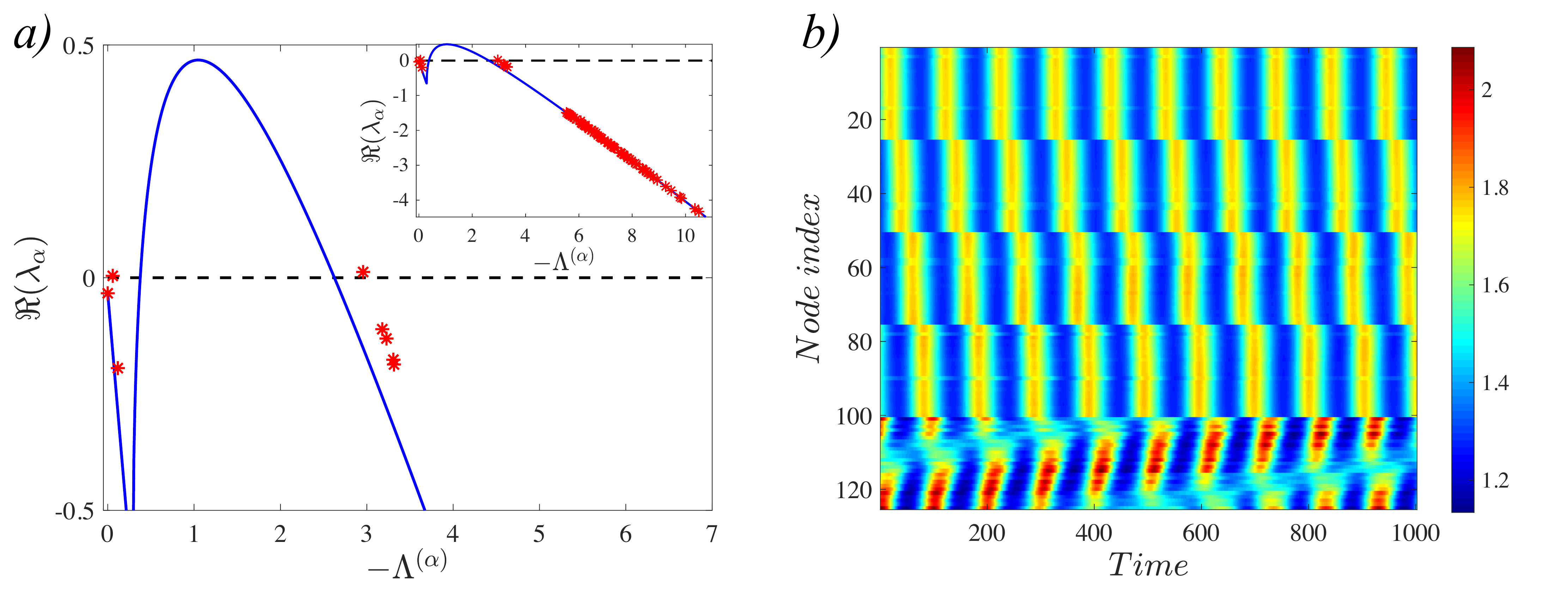}
    \caption{\textbf{Chimera states in directed modular networks} We have used the Brusselator model with parameters $a=1.5$, $b=3.2$, $c=d=1$, $D_u=0.6$ and $D_v=3.75$. $\textbf{a)}$ The dispersion relation shows that a modular and a non-modular eigenvalue are simultaneously unstable. In the inset are given the real and the imaginary part of the dispersion relation.  \textbf{b)} The chimera states evolution is shown as a function of time where the last module is incoherent compared to the remaining four ones. } 
    \label{fig:direct_chimera}
\end{figure*}

To proceed with the stability analysis, we linearise the reaction-diffusion system (\ref{eq:RD}) around the uniform steady state $(\textbf{u}^*, \textbf{v}^*)$:
\begin{eqnarray*}
\begin{pmatrix} \dot{\delta \textbf{u}} \\ \dot{\delta \textbf{v}} \end{pmatrix}&=&\begin{pmatrix}
f_u \boldsymbol{\mathcal{I}} + D_u \boldsymbol{\mathcal{L}} & f_v \boldsymbol{\mathcal{I}}\\g_u \boldsymbol{\mathcal{I}} & g_v + D_v \boldsymbol{\mathcal{L}} \end{pmatrix}\begin{pmatrix} \delta \textbf{u} \\ \delta \textbf{v}\end{pmatrix}\\&=& \big(\boldsymbol{\mathcal{J}} + \boldsymbol{\mathcal{D}}\big)\begin{pmatrix} \delta \textbf{u} \\ \delta \textbf{v}\end{pmatrix}
\end{eqnarray*}
where here $\delta \textbf{u}$, $\delta \textbf{v}$ are the vectors of the perturbations. The Jacobian matrix for the reaction terms, evaluated at the equilibrium point is given by  $\boldsymbol{\mathcal{J}}=\begin{pmatrix}
f_u \boldsymbol{\mathcal{I}} & f_v \boldsymbol{\mathcal{I}}\\g_u \boldsymbol{\mathcal{I}} & g_v \boldsymbol{\mathcal{I}} \end{pmatrix}$, and the diffusion one for both species is $\boldsymbol{\mathcal{D}}=\begin{pmatrix}
D_u  \boldsymbol{\mathcal{L}} & \boldsymbol{\mathcal{O}}\\\boldsymbol{\mathcal{O}} & D_v \boldsymbol{\mathcal{L}} \end{pmatrix}$, where $\boldsymbol{\mathcal{O}}$ is the $N\times N$ zero-valued matrix. We consider the basis of the eigenvectors $\Phi_i^{\alpha}$, with $\alpha=1,\dots, N$, of the Laplacian operator $ \boldsymbol{\mathcal{L}}$ to the corresponding eigenvalues $\Lambda^{\alpha}$. Due to the asymmetry of the adjacency matrix $\mathbf{A}$, the spectrum $\Lambda^{\alpha}$ is, in principle, complex. In particular is has been shown in \cite{natcomm, epjb} that for balanced directed networks where the numbers of incoming edges is equal to the outgoing ones $k_i^{out}=k_i^{in}$, such a basis always exists. This will be also the case in the following discussion. In order to solve this linearised system, we expand the perturbations on the basis of the eigenvectors, i.e. $\delta u_i= \sum_{\alpha=1}^N b_{\alpha} \Phi_i^{\alpha}$ and $\delta v_i= \sum_{\alpha=1}^N  c_\alpha \Phi_i^{\alpha}$. Following this, it is possible to reduces the $2N \times 2N$ system to a $2\times 2$ eigenvalue problem, for each value of the node index $\alpha=1,\dots, N$: 
\begin{eqnarray*}
\det\begin{pmatrix} \textbf{J}_\alpha - \lambda_\alpha\textbf{I}\end{pmatrix}&=&\det\begin{pmatrix}
f_u + D_u\Lambda^{\alpha}-\lambda_\alpha \hspace*{-.5cm}& f_v \\g_u  & g_v + D_v\Lambda^{\alpha}-\lambda_\alpha \end{pmatrix}\\&=&0
\end{eqnarray*}
where $\textbf{J}_\alpha$ is the $2\times 2$ extended Jacobian, {i.e.} $\textbf{J}_\alpha\equiv\textbf{J} + \textbf{D}\boldsymbol{\Lambda}^{\alpha}$ where $\textbf{D}=diag(D_u,D_v)$, $\boldsymbol{\Lambda}^{\alpha}=diag(\Lambda^{\alpha},\Lambda^{\alpha})$. If the real part of $\lambda_\alpha$, known as the dispersion relation, has at least a mode $\Lambda^{\alpha\neq 0}$ for which takes positive values, then the steady state becomes unstable. Mathematically this is formalised as:
\begin{equation}
\lambda_\alpha=\frac{1}{2}\left[\left(\mathrm{tr}\textbf{J}_\alpha\right)_{Re}+\gamma\right] + \frac{1}{2}\left[\left(\mathrm{tr}\textbf{J}_\alpha\right)_{Im}+\delta\right]\iota
\label{eq:disp_rel}
\end{equation}
where 
\begin{eqnarray*}
\gamma&=&\sqrt{\frac{A+\sqrt{A^2+B^2}}{2}}\\
\delta&=&\mathrm{sgn}(B)\sqrt{\frac{-A+\sqrt{A^2+B^2}}{2}}\\
A&=&\left[\left(\mathrm{tr}\textbf{J}_\alpha\right)_{Re}\right]^2 - \left[\left(\mathrm{tr}\textbf{J}_\alpha\right)_{Im}\right]^2-\left[\left(\det\textbf{J}_\alpha\right)_{Re}\right]^2\\
B&=&2\left(\mathrm{tr}\textbf{J}_\alpha\right)_{Re}\left(\mathrm{tr}\textbf{J}_\alpha\right)_{Im}-\left[\left(\det\textbf{J}_\alpha\right)_{Im}\right]^2.
\end{eqnarray*}
In the above expression we have denoted by $\mathrm{sgn}(\cdot)$ the sign function and $()_{Re}$, $()_{Im}$ indicate, the real and imaginary parts of the respective function. 
To proceed further with the calculations, we make use of the definition of a square root of a complex number. By taking $z=a+b\iota$ 
{where $\iota=\sqrt{-1}$ is the imaginary unit}, then
\begin{equation*}
\sqrt{z}=\pm \left(\sqrt{\frac{a+|z|}{2}}+ \mathrm{sgn}(b)\sqrt{\frac{-a+|z|}{2}}\iota\right).
\end{equation*}
The instability sets in when 
$|\left(\mathrm{tr}\textbf{J}_\alpha\right)_{Re}|\leq \gamma$, a condition that can be expressed by the following inequality \cite{natcomm}: 
\begin{equation}
S_2(\Lambda_{Re}^{(\alpha)})[\Lambda_{Im}^{(\alpha)}]^2\leq -S_1(\Lambda_{Re}^{(\alpha)}),
\label{eq:instdom}
\end{equation}
where $(\Lambda_{Re}, \Lambda_{Im})$ span the complex plane where lie the eigenvalues of the Laplacian operator. The explicit form
of the polynomials $S_1$, $S_2$ is described by:
\begin{eqnarray*}
S_1(x)&=& C_{14}x^4+C_{13}x^3 + C_{12}x^2 + C_{11}x+ C_{10}\\
S_2(x)&=& C_{22}x^2 + C_{21}x+ C_{20}.
\label{eq:s1s2funct}
\end{eqnarray*}
The above constants are given by:
\begin{eqnarray*}
C_{14}&=&\sigma\left(1+\sigma\right)^2\\
C_{13}&=&\left(1+\sigma\right)^2\left(\sigma J_{11}+J_{22}\right)+2\mathrm{tr}\textbf{J}\sigma\left(1+\sigma\right)\\
C_{12}&=&\det\textbf{J}\left(1+\sigma\right)^2 + \left(\mathrm{tr}\textbf{J}\right)^2\sigma + 2\mathrm{tr}\textbf{J}\left(1+\sigma\right)\left(\sigma J_{11}+J_{22}\right)\\
C_{11}&=&2\mathrm{tr}\textbf{J}\left(1+\sigma\right)^2\det \textbf{J} + \left(\mathrm{tr}\textbf{J}\right)^2\left(\sigma J_{11}+J_{22}\right)\\
C_{10}&=&\det \textbf{J}\left(\mathrm{tr}\textbf{J}\right)^2\\
C_{22}&=&\sigma\left(1- \sigma\right)^2\\
C_{21}&=&\left(\sigma J_{11}+J_{22}\right)\left(1- \sigma\right)^2\\
C_{20}&=&J_{11}J_{22}\left(1- \sigma\right)^2.
\label{eq:c2coeff}
\end{eqnarray*}
The above system undergoes a Turing-Hopf instability, induced by the (directed) topology of the interaction network. Such mechanism responsible for the emergence of pattern in directed networks, is known as the topology-driven instability \cite{natcomm}. In Fig. \ref{fig:direct_sync} we show the cluster synchronisation phenomenon in a directed modular network where the pattern are triggered by the directedness of the underlying network and the oscillations are as a consequence of a complex spectrum of the Laplacian operator. If apart the unstable modular mode, we could chose also to destabilise one of the non-modular ones then chimera states would appear in this case as shown in Fig. \ref{fig:direct_chimera}.\\

\onecolumngrid


\twocolumngrid

\end{document}